\documentclass[conference]{IEEEtran}
\IEEEoverridecommandlockouts
\usepackage{cite}
\usepackage{amsmath,amssymb,amsfonts}
\usepackage{algorithmic}
\usepackage{graphicx}
\usepackage{textcomp}
\renewcommand\IEEEkeywordsname{Keywords}
\def\BibTeX{{\rm B\kern-.05em{\sc i\kern-.025em b}\kern-.08em
    T\kern-.1667em\lower.7ex\hbox{E}\kern-.125emX}}
\begin{document}

\title{Estimation of Optical Aberrations in 3D Microscopic Bioimages \\
\thanks{
{$^{\mathrm{1}}$}https://github.com/kiraving/aberration}
}
\author{\IEEEauthorblockN{
Kira Vinogradova} %
\IEEEauthorblockA{\textit{Center for Systems Biology Dresden} \\
\textit{Max Planck Institute of Molecular Cell Biology and Genetics}\\
Dresden, Germany \\
vinograd@mpi-cbg.de}
\and
\IEEEauthorblockN{
Eugene W. Myers} %
\IEEEauthorblockA{\textit{Center for Systems Biology Dresden} \\
\textit{Max Planck Institute of Molecular Cell Biology and Genetics}\\
Dresden, Germany \\
myers@mpi-cbg.de}
}

\maketitle

\begin{abstract}

The quality of microscopy images often suffer from optical aberrations. 
These aberrations and their associated point spread functions have to be quantitatively estimated to restore aberrated images. 
The recent state-of-the-art method PhaseNet, based on a convolutional neural network, can quantify aberrations accurately but is limited to images of point light sources, e.g. fluorescent beads. 
In this research, we describe an extension of PhaseNet enabling its use on 3D images of biological samples. 
To this end, our method incorporates object-specific information into the simulated images used for training the network.
Further, we add a Python-based 
restoration of images via Richardson–Lucy deconvolution. We demonstrate that the deconvolution with the predicted PSF can not only remove the simulated aberrations but also improve the quality of the real raw microscopic images with unknown residual PSF. 
We provide code{$^{\mathrm{1}}$} for fast and convenient prediction and correction of aberrations.
\end{abstract}


\section{Introduction}
An aberration is a deviation of a light beam from the trajectory described by geometrical optics. Optical aberrations can be chromatic (color aberrations) or monochromatic. Monochromatic aberrations make images look blurred or distorted. In adaptive optics (AO), the branch of microscopy that studies estimation and correction of aberrations, an estimation of monochromatic aberrations is possible via finding the presence and amplitudes of the Zernike polynomials \cite{von1934beugungstheorie} describing the PSF at a point in space. In terms of Deep Learning, estimation of aberrations means predicting a 1D vector of amplitudes for each Zernike mode.

AO methods aim at enhancing the image quality by correcting optical aberrations, which need to be estimated first. There are two ways to estimate aberrations: using a wavefront sensor \cite{dms} or computationally, where the latter mostly relies on optimization algorithms \cite{vishniakou2021differentiable,optimisation} or deep learning \cite{saha2020practical,hu2022efficient,whang2021zernike,dnn,Tian19}. The advantage of sensorless  approaches based on deep learning is their simplified design of the optical system; however, they require a large number of images for training.

State-of-the-art computational methods are limited to either trivial point light sources \cite{saha2020practical,vishniakou2021differentiable,whang2021zernike,dnn,Tian19}, which reflect the point spread function (PSF) by definition, or to objects with a simple well-defined shape like microtubules \cite{hu2022efficient}. The work \cite{Tian19} provides a comparison of six models, with five of them being Convolutional Neural Networks (CNN) of different depths. Remarkably, the authors demonstrated that the best performance on simulated images of fluorescent beads was achieved by a rather small network with 7 convolutional layers combined with 5 max pooling and 3 dense layers. Some of the state-of-the-art methods are designed only for two-dimentional images \cite{dnn,Tian19}, although the PSF affecting volumetric objects has three dimentions.

To the best of our knowledge, the PhaseNet publication \cite{saha2020practical} provides the only publicly available 
datasets with experimentally introduced aberrations via a point-scanning AO microscope with a spatial light modulator and a custom-built widefield AO microscope with a deformable mirror. The PhaseNet datasets are aberrated images of fluorescent beads that have a trivial spherical shape.

In this research, we explore the applicability of PhaseNet to objects of a complex shape — the protoplasmic astrocytes in the hippocampus of Rattus norvegicus \cite{bushong2004maturation} — for which we synthetically generate aberrated images.

\section{Methods}

\subsection{PhaseNet}

The core method used in this study, PhaseNet \cite{saha2020practical}, is a CNN that takes a 3D image as an input and outputs the amplitudes for each Zernike mode that it was trained on. The CNN consists of five convolutional blocks (two 3D convolutional layers and one 2D max-pooling layer) and three dense layers. 

PhaseNet has been trained on synthetic images of 3D aberrated points. The synthetic generator produced random aberrations with the amplitudes in a predefined range. 
The training and validation datasets were formed using one generator. PhaseNet has been tested on randomized synthetic data from the same generator and on experimentally acquired 3D images of fluorescent beads on two AO microscopes. 

The model predicted the amplitudes for 11 nontrivial aberration modes, and the visualization of the PSF and the wavefront reconstruction were shown. The Python code for the model and data generation along with the data have been made publicly available on Github by the authors. For details regarding the data acquisition and CNN training, please refer to the original publication \cite{saha2020practical}.

\begin{figure}[htb]
\centerline{\includegraphics[width=\linewidth]{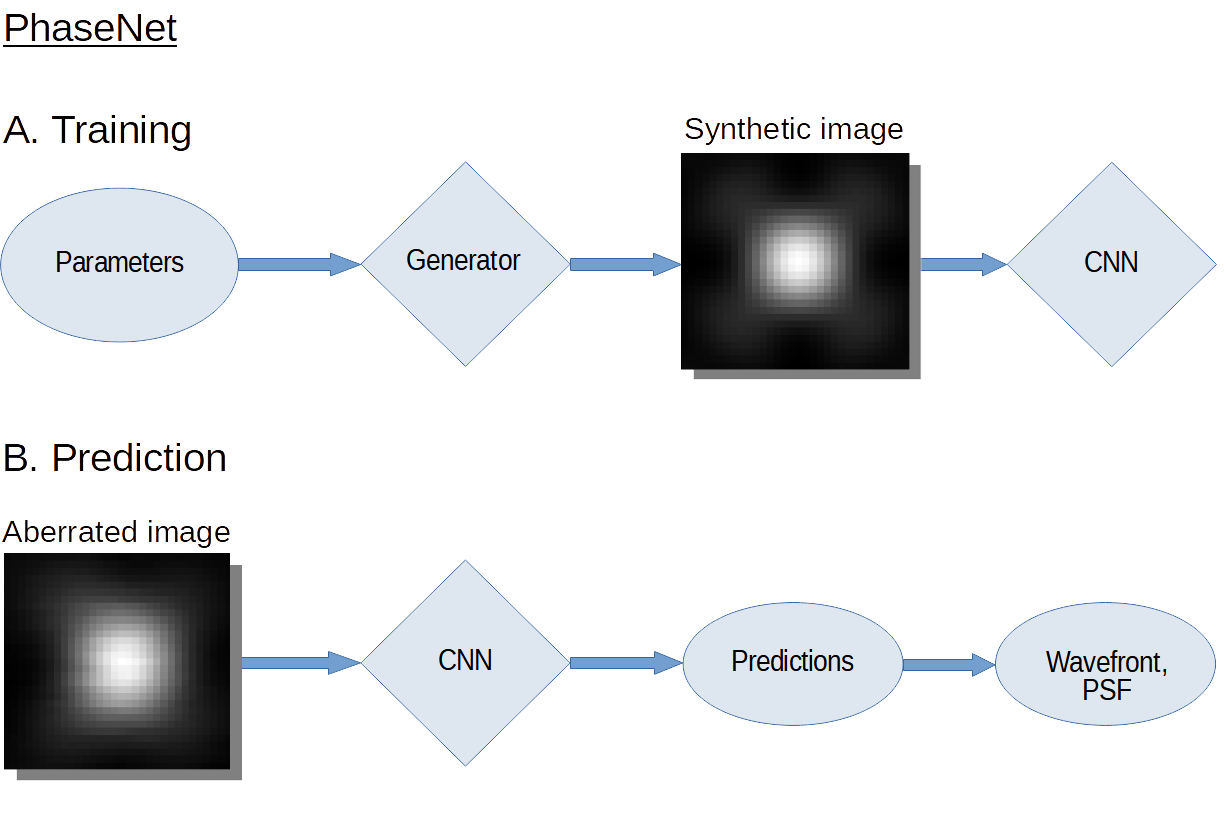}}
\caption{Schematic of PhaseNet \cite{saha2020practical} algorithm. 
Training: the generator takes the input parameters of the microscopy system (refractive index, numerical aperture, etc.) and parameters for data generation (size of the point source, noise to add, etc.) and produces the 3D PSF together with the ground truth (aberration modes and their amplitudes). The CNN is trained in a supervised manner on the generator's output. Prediction: the trained CNN predicts the vector of the aberration modes with their amplitudes (one float number per mode) for an input 3D image of a bead. 
The predictions are used to visualize the PSF and the wavefront.
}
\label{fig1}
\end{figure}

Fig.~\ref{fig1} summarizes the training procedure and usage of PhaseNet.

\subsection{Synthetic Data Generator}\label{SYN}
The synthetic generator in PhaseNet produces random PSFs. The user defines the nomenclature (ANSI and Noll are available), the list of aberration modes, the range of amplitudes, the size of the generated PSF, and the output size (the size of a crop of the PSF). The PSF can be cropped, and the cropping position can be configured with the parameters “jitter” and “max jitter”. If “jitter” is “false” (default), the crop is made in the center of the image; if it is “true”, the crop is taken from a random part of the image. 
“Max jitter” defines how far from the center a crop can be taken (by default, anywhere in the image). The user can select the $z$ planes of the 3D image to train the network on (for example, only a couple of central planes). It has been demonstrated \cite{saha2020practical} that the best results are achieved when all $z$ planes are used, which means setting “max jitter” to “None” (default) and training on full-sized 3D images.

The parameters of the optical system (refractive index, light wavelength {$\lambda$}, numerical aperture of the detection objective (NA), voxel size) must be provided for correct calculation of the synthetic PSFs. The generator convolves a point (sphere) 
with the synthetic PSF. Optionally, Gaussian noise can be added after the convolution if its parameters (standard deviation, mean, and signal-to-noise ratio (SNR)) are provided. For each of these parameters, either a single float value 
or a range of values are accepted. Also, Gaussian blur can be added.

\subsection{Noise Parameters}\label{NP}
PhaseNet takes dataset-specific parameters, such as the noise parameters, which are defined by the acquired images.

To retrieve the noise parameters, we used the “measure” function in Fiji software \cite{schindelin2012fiji} on selected areas of the image. The bright parts of the image belonging to the astrocyte were considered the foreground, the dark areas — background (noise). The generator accepts a range for each noise parameter, therefore the mean intensity was measured on 
different parts of the astrocyte, as well as the mean
and standard deviation of the noise were measured twice. 
By definition, the SNR is the ratio of the average power of a signal to the average power of a background noise. The SNR was calculated as the foreground mean divided by the noise mean.

\subsection{
Data Generation with Phantoms}\label{GEN}
To include the information about the object structure, we passed a crop of the raw unaberrated image into the generator, where 
it was convolved using synthetic PSFs. 
In our experiments with synthetic data, the unaberrated image is simply the downloaded raw image. In an experimental setup with acquisition of real data using an AO microscope, an unaberrated image would be an image without aberrations introduced by a deformable mirror (or by another hardware, e.g. a spatial light modulator).
It is important to provide such an unaberrated image to ensure that the only PSF present is the residual PSF of the microscope because it is present in all the data coming from this microscope, whereas the introduced aberrations appear in images only when the AO microscope is instructed to add them.

Our approach for generating images using an actual image aims to add prior information about the object shape, the residual PSF of the microscope, and other noise coming from the microscope besides Gaussian noise, which the generator synthesizes.

We generated synthetically aberrated 3D images of astrocytes by cropping the input image inside the generator, convolving the crop with random synthetic PSFs, and adding random Gaussian noise. We refer to the input image that was passed into the generator as a “phantom”. Training with two or more images passed as phantoms is named “training with multiple phantoms” in this paper.

Generation of the images simulated the acquisition of a dataset with 11 Zernike modes. 
We limited our dataset to the Zernike polynomials up to the fourth order because higher-order polynomials have a low impact on the image quality. The trivial modes (piston, x-tilt, y-tilt, and defocus) were excluded, as has been done in the original PhaseNet project.

In this research, we investigate whether the training with a phantom or multiple phantoms is necessary for application of PhaseNet to complex objects.

\subsection{Restoration via Deconvolution}\label{AA}
The goal of predicting aberrations is to precisely restore distorted images. Restoration can be achieved using deconvolution, an operation inverse to convolution, whose purpose is to restore a convolved image to its state before the convolutional kernel was applied. In our case, the convolutional kernels are PSFs (the synthetically introduced PSFs and the residual PSF of the microscope), therefore deconvolution with predicted PSFs can be used for restoring the images. 

In our workflow, we incorporate the Python-based library Flowdec \cite{Czech460980} which implements the 3D deconvolution using the Richardson–Lucy algorithm \cite{richardson1972bayesian,lucy1974iterative} for restoration of aberrated images. Compared to commonly used DeconvolutionLab2 \cite{sage2017deconvolutionlab2}, Flowdec has achieved the results faster. The researchers provided a Jupyter notebook on their GitHub page comparing Flowdec with DeconvolutionLab2 on a computer with 64G RAM, Intel Xeon CPU (x2), and Nvidia GTX 1080 GPU (x2), which demonstrated that Flowdec was 55 times faster on average.

We chose Flowdec because it is fast, able to take the PSFs generated by our synthetic generator, and can be called directly from Python scripts or Jupyter notebooks.

\begin{figure}[htb]
\centerline{\includegraphics[width=\linewidth]{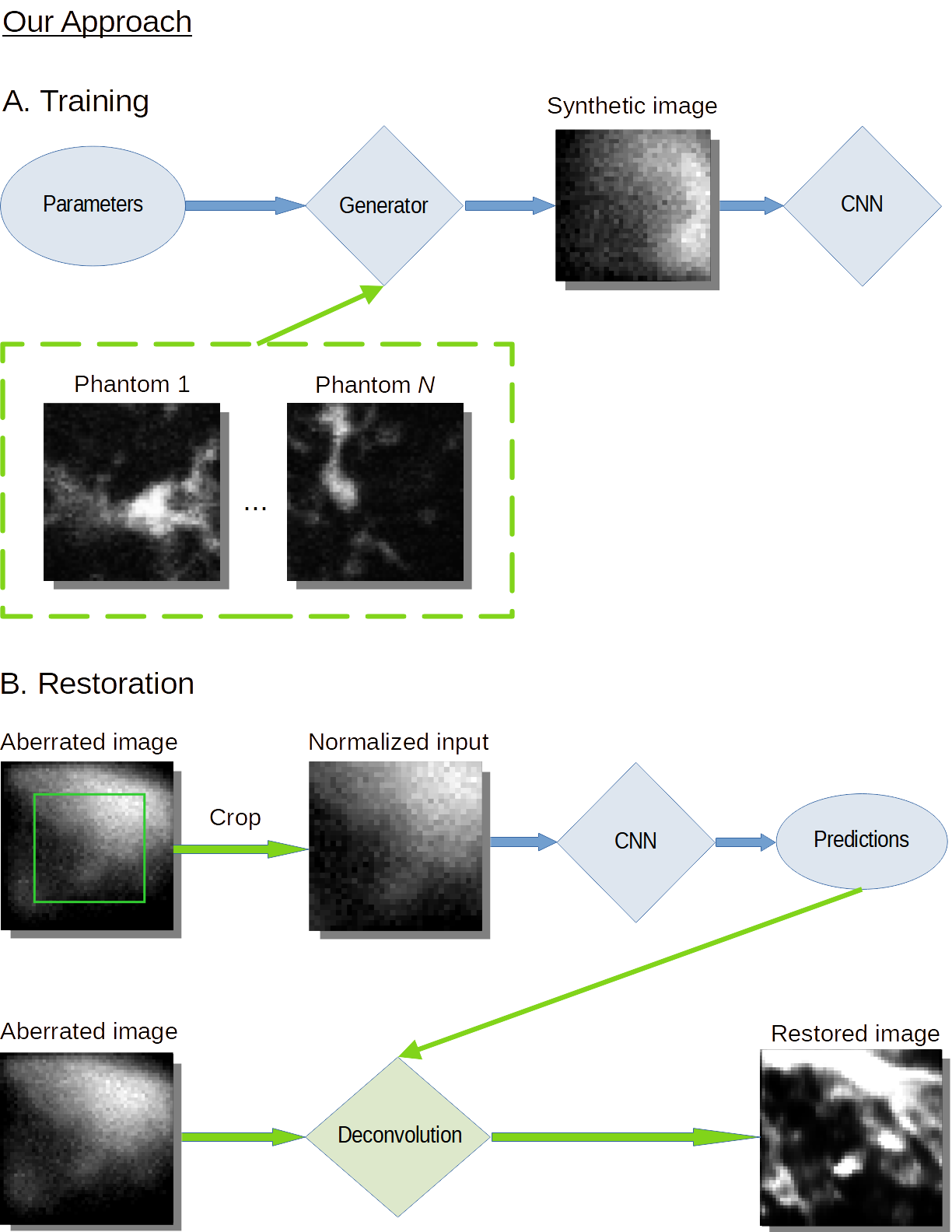}} 
\caption{Schematic of training on phantoms followed by image restoration.
Training: the generator takes the input parameters of the microscopy system, parameters for data generation, and phantoms (unaberrated 3D images of the study samples). At each iteration, it convolves a PSF with a random 3D crop of a randomly chosen phantom and outputs an aberrated image with the ground truth, which are passed to the CNN for supervised training. Restoration: the trained CNN predicts the vector of the aberration modes with their amplitudes for an aberrated image of the sample. The CNN predictions can be used for generating the PSF, which can then be used as a deconvolution kernel in restoration. The restoration was done using the Flowdec library in Python. The green parts of the scheme denote our modifications of PhaseNet.}
\label{fig3}
\end{figure}

Our approach of training on phantoms and restoring images using Flowdec is schematically shown in Fig.~\ref{fig3}.

\section{Data}
\subsection{Source Images}

For this research, 3D images with known characteristics of the microscopy setup (refractive index, {$\lambda$}, NA, voxel size) were 
taken from the open database Cell Image Library (cellimagelibrary.org). The images were of rat hippocampus at different ages taken at various resolutions \cite{bushong2004maturation}. We chose two images \cite{ly17,ly22} on which the voxel size, fluorescent dye, and developmental stage were identical. Further, we refer to them as the first \cite{ly17} and the second \cite{ly22} raw image or dataset.

The images of astrocytes in a 4-week-old rat hippocampal area CA1 intracellularly injected with Lucifer Yellow \cite{belichenko1995studies} (${\lambda}=488$~nm) were acquired with a single photon confocal microscope with $\mathrm{NA}=1.4$, oil immersion medium, and voxel size of (0.068519, 0.068519, 0.2) ${\mu}$m in $(x, y, z)$. For more details, please refer to \cite{bushong2004maturation,ly17,ly22}.

\subsection{
Training and Test Sets}
\begin{figure}[htb]
\centerline{\includegraphics[width=\linewidth]{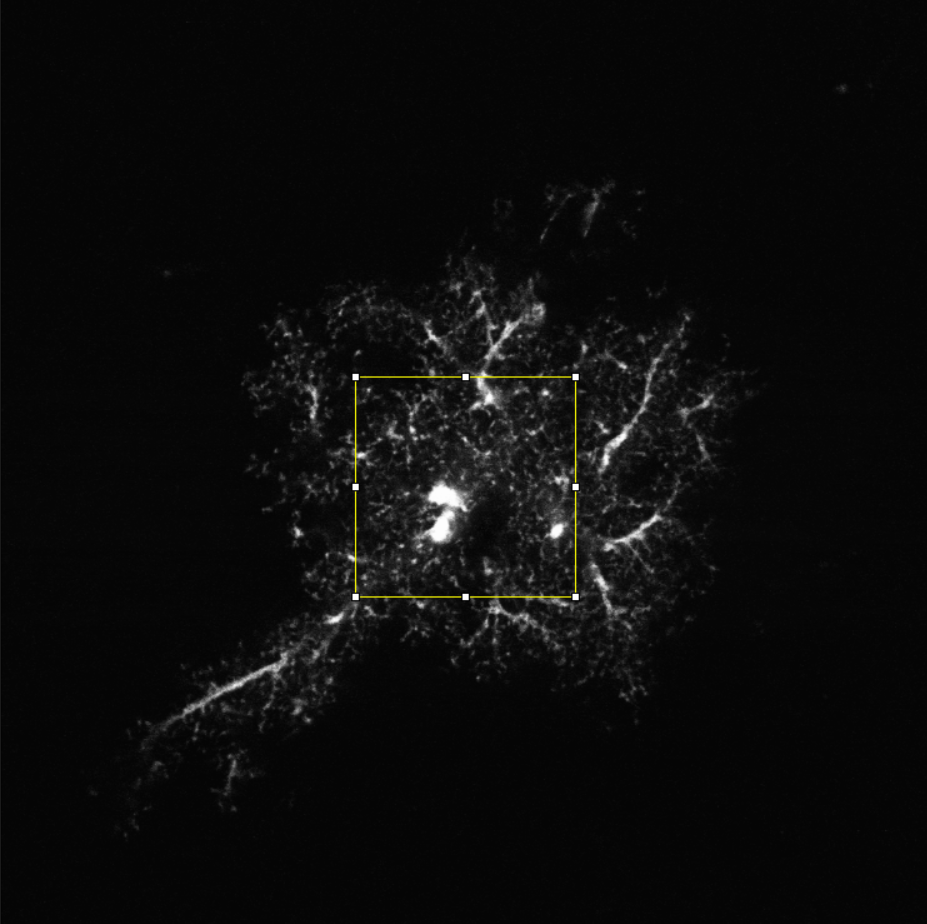}}
\caption{Cropping out the object in the 3D image \cite{ly22} (the central 2D slice shown in Fiji).}
\label{fig4}
\end{figure}

The astrocyte images had a large size, and objects were not present in all parts of the images. To ensure that the input images for PhaseNet contained an object, a smaller 3D area was cropped out in both raw images. A 2D slice of the second image \cite{ly22} with a manual cropping is shown in Fig.~\ref{fig4}. To speed up the training, even smaller crops of (32, 32, 32) pixels (as in \cite{saha2020practical}) were made, in the center or off-center, before they were passed to the CNN.

To the best of our knowledge, only images of fluorescent beads with aberrations experimentally introduced by AO microscopes were publicly available as the time of preparation of the manuscript. Acquisition of bioimages of such kind is time-consuming and expensive because it requires an AO microscope in the first place, as well as a suitable fixed fluorescent biosample. Due to unavailability of bioimages with experimentally introduced aberrations for our research, we synthesized different randomized training, validation, and test data using our generation approach described in Section \ref{GEN} by passing the downloaded images of astrocytes as phantoms to the generator. The generator was able to synthesize realistic images of fluorescent beads according to our comparison of the synthetic images to the real ones available in the publication \cite{saha2020practical}. Based on this, we assumed that the generator could simulate realistic aberrations.

Images in the training and validation sets had random amplitudes for each of 11 aberration modes. The test sets consisted of 11 series of images for each aberration mode in order to make the visualization of the predictions more convenient and to compare them with the plots from the original PhaseNet paper. The test images in the series for each aberration mode were convolved with synthetic PSFs, which had only a random amplitude of the target mode whereas other amplitudes were not present.

In addition to our synthetic data generation approach, we modified the training procedure so that for validation the network can be provided with a data generator that is different from the train set generator, or supplied with an array of images. The latter functionality can be used in other research, where real experimental images are available, for detection of models' overfitting to synthetic train data.  

\section{Results}
\subsection{Applying PhaseNet}
We retrained PhaseNet with the new parameters of the microscope that determine the synthetic PSFs. The network expects a certain pixel size, therefore the pretrained weights from the original PhaseNet study cannot be reused for data with objects of a different scale. In addition, in the training configuration, we changed the noise parameters in the data generator according to the microscope setup (Section \ref{NP}). The model was trained on synthetic PSFs (point-like objects) without any other modifications except for 
changing the microscope's and the noise parameters, and the pixel size. 
The training procedure is shown in Fig.~\ref{fig1}A.

Then, synthetically aberrated images of astrocytes were generated by cropping the image and convolving the crop with random synthetic PSFs (Section \ref{GEN}). Generation of the images simulated the acquisition of the dataset consisting of 11 test series of images for each aberration mode. We tested retrained PhaseNet on this new data and evaluated the predictions using the mean squared error (MSE) averaged across all predictions. The model was unable to predict aberrations, resulting in the MSE = %
0.001718 on random crops. A possible reason could be overfitting on the training object shape: the network learned how aberrated point sources look like but could not generalize to nonspherical objects in the test set. Alternatively, the cause could be the inability of PhaseNet to understand the aberrations on complex shapes, which is investigated below.

\begin{figure}[htb]
\centerline{\includegraphics[width=\linewidth]{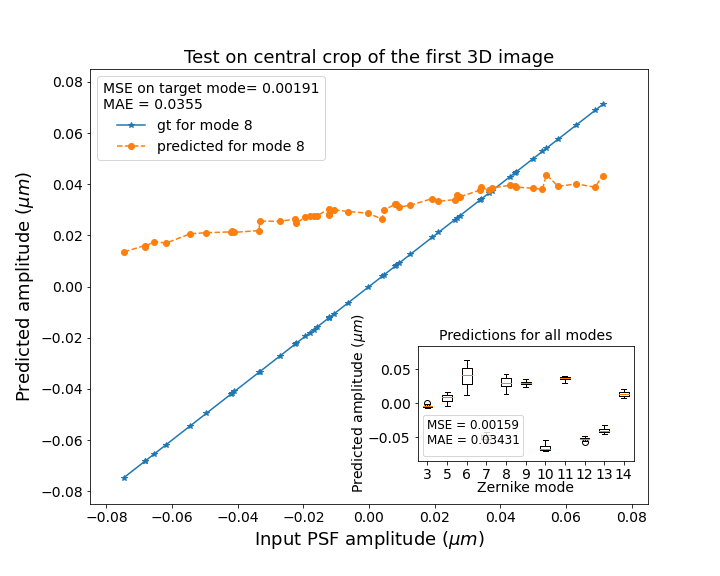}}
\caption{Results of PhaseNet trained on points and tested on astrocytes. Blue stars on the diagonal line are the desired ground truth predictions for the present aberration mode (mode 8, horizontal coma). Yellow dots indicate the predicted values. The inset shows the statistics of predictions for all the modes.}
\label{fig5}
\end{figure}

For visualization purposes, the results for one of the aberration modes (mode 8 in ANSI nomenclature, horizontal coma) are plotted in Fig.~\ref{fig5}. 
In the plot, blue stars on the diagonal line are the desired ground truth predictions for the present aberration mode, yellow dots are the predicted values. The inset in the bottom right corner shows statistics of predictions for all modes (minimum, maximum, median, the first and third quartiles, and outliers). Ideally, the variation of predictions for the modes, except the present one, should be minimal and centered around zero {$\mu$}m.

\subsection{Training with Phantom Image}
\begin{figure}[htb]
\centerline{\includegraphics[width=\linewidth]{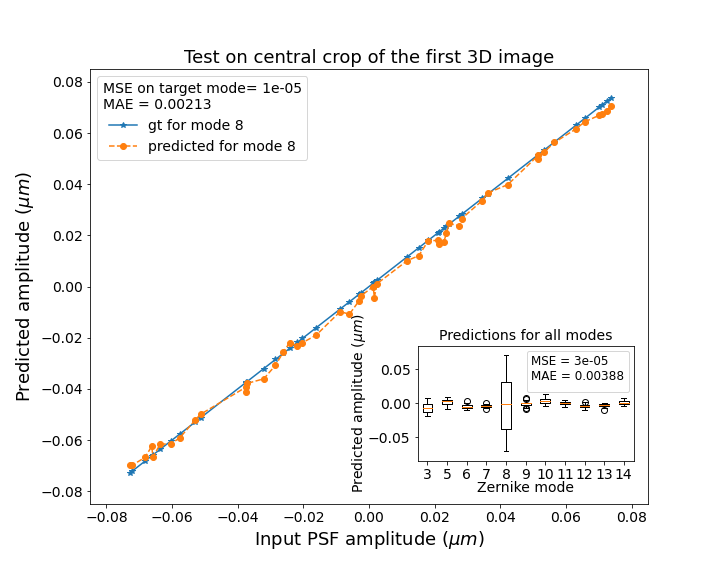}}
\caption{Results of training on the phantom with central cropping and testing on the central crop convolved with random PSFs. Blue stars on the diagonal line are the desired ground truth predictions for the present aberration mode (mode 8, horizontal coma). Yellow dots indicate the predicted values. The inset shows the statistics of predictions for all modes.}
\label{fig6}
\end{figure}

To investigate whether PhaseNet can learn the aberrations on an object of a nontrivial shape, we trained the model with the image \cite{ly17} as the phantom. In this experiment, the generator was instructed to crop the center of the phantom, meaning that the object on the images remained the same and only the introduced aberrations varied. The results (Fig.~\ref{fig6}) proved that the model can learn the representations of aberrations on a nontrivial object, such as a fragment of an astrocyte, if the training data includes the information about the object.

To understand whether the model overfits on the object that was present in the supplied phantom, the model was tested on the random crops (“jitter” = “true”, see Section \ref{SYN}) from the first image \cite{ly17} around the center and from the second image \cite{ly22} (unseen by the model during training). A significant drop in performance from MSE = 0.000027 on the central crop to 0.000299 on the random crops around the center and 0.002186 on the unseen data indicated that the model strongly overfits on the object if its shape does not vary.

\subsection{Training with Phantom and Random Cropping}
\begin{figure}[htb]
\centerline{\includegraphics[width=\linewidth]{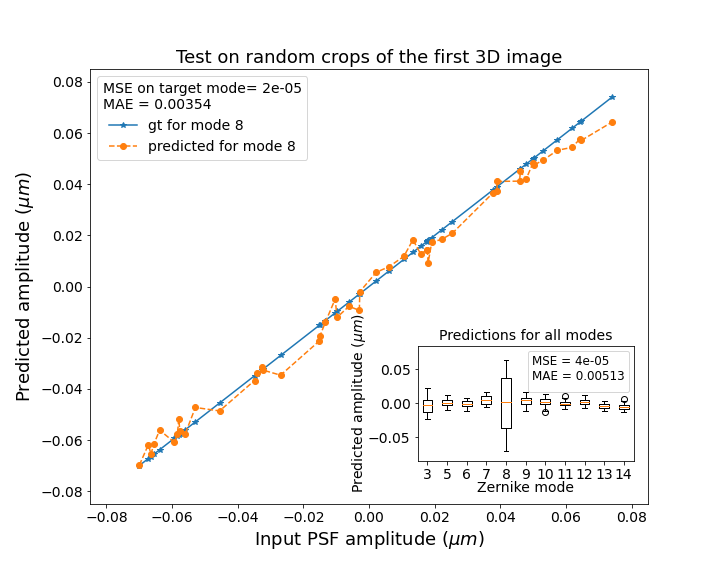}}
\caption{Results of the training on random crops of the phantom and testing on random crops. Blue stars on the diagonal line are the desired ground truth predictions for the present aberration mode (mode 8, horizontal coma). Yellow dots indicate the predicted values. The inset shows the statistics of predictions for all modes.}
\label{fig7}
\end{figure}

We explored whether the model can learn the aberrations if the shape of the object varies. The variety of the training data was enlarged by the random off-center cropping (“jitter” = “true”, see Section \ref{SYN}).
This experiment proves (Fig.~\ref{fig7}) that the model can perform well on objects of varying shapes if it has seen these shapes during the training (Table~\ref{tab1}, row “Random crop").

\subsection{Training with Multiple Phantoms}
We supplied the generator with two phantoms: one used in the previous experiments and another cropped from the second image. Our generator randomly selected a phantom from the list of images at each iteration of generating a synthetically aberrated image (at each training step). The phantoms were randomly cropped. Testing on random crops from the first and second images with random aberrations gave almost the same MSE on both sets: 0.000129 on the first one and 0.000131 on the second. Clearly, this model performed better on the second dataset than the previous CNNs because it had seen the objects from this dataset during the training. 

To further prove the idea that the model overfits on the shapes of the objects from the training set, we tested it on another region from the second image that did not overlap with the crops used for training and validation. As a result, we observed a significant drop in the MSE to 0.001554. Nevertheless, training on more various data led to a lower MSE compared to previous experiments.

Additionally, we tried training on the central region of the first image, validation on the central region of the second image, and testing on another region of the second image. Overfitting was observed after 400 epochs, which supports our assumption that the network overfits on the seen shapes.

The results of all experiments are summarized in Table~\ref{tab1}. The models were trained for 50,000 epochs with 5 steps per epoch and the batch size of 2 images. Training of one model on a laptop with Intel Core i7-8850H CPU, single NVIDIA Quadro P3200 GPU with 8 Gb memory, and 64 Gb RAM lasted 30 hours. 

\begin{table}[h!] 
\caption{Comparison of results}
\begin{center}
\begin{tabular}{|c|c|c|c|}
\hline
\textbf{ }&\multicolumn{3}{|c|}{\textbf{Mean MSE of testing on:}} \\
\cline{2-4} 
\textbf{Training on} & \textbf{\textit{Central crop}}& \textbf{\textit{Random crop}}& \textbf{\textit{Unseen set}} \\
\hline
Points (III-A)& na{$^{\mathrm{a}}$} & na{$^{\mathrm{a}}$} & 0.001718 \\

\hline
Central crop (III-B)&  0.000027& 0.000299 & 0.002186 \\
\hline
Random crop (III-C)& 0.000040& 0.000045 & 0.001793  \\
\hline
Two phantoms{$^{\mathrm{b}}$} (III-D)& 0.000090& 0.000129 & 0.001554 \\
\hline
\multicolumn{4}{l}{$^{\mathrm{a}}$Sets with central and random crops were unseen.}
\\
\multicolumn{4}{l}{$^{\mathrm{b}}$Training with multiple phantoms, two were provided.}
\end{tabular}
\label{tab1}
\end{center}
\end{table}

\subsection{Restoration of the Aberrated Images}
\begin{figure}[htb]
\centerline{\includegraphics[width=\linewidth]{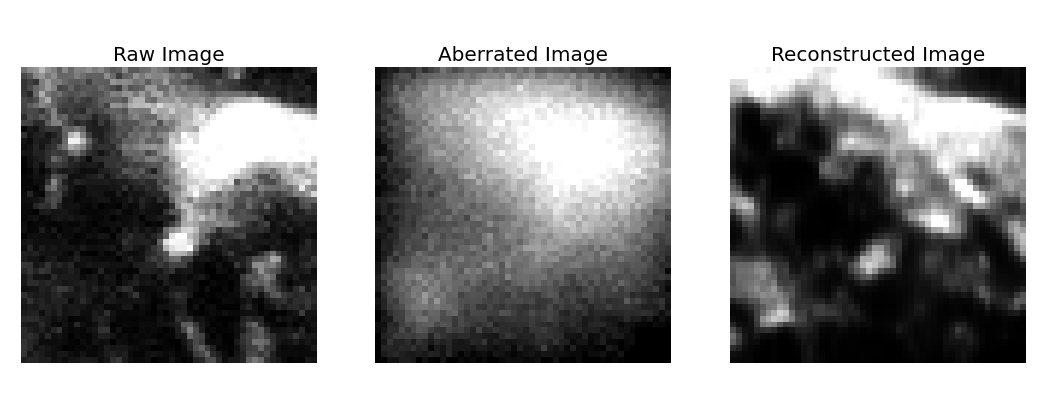}}
\caption{Restoration of an aberrated 3D image using deconvolution with the predicted 3D PSF. 2D slices are shown.}
\label{restor}
\end{figure}

As shown in Fig.~\ref{fig3}, our workflow includes the restoration of aberrated 3D images. An aberrated image is cropped in three dimensions and passed to the model which predicts the aberration modes with their amplitudes. The predictions are used for generating the PSF. The full-size aberrated 3D input image is then deconvolved with the PSF image. An example of a restoration of the synthetically aberrated image is shown in Fig.~\ref{restor}.

\subsection{Restoration of the Source Images}
It is possible to improve the quality of the initial source images by the deconvolution of the initial raw image with the PSF predicted by our model trained on the image as the phantom. To do so, the raw image has to be cropped to the size accepted by the network. The network takes the crop without additional modifications and outputs the vector of amplitudes for the residual PSF. The PSF is generated based on the prediction and used as the deconvolutional kernel. 

\begin{figure}[htb]
\centerline{\includegraphics[width=\linewidth]{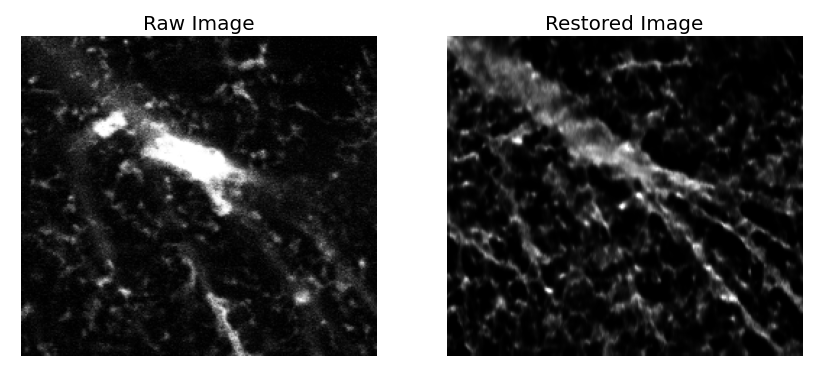}}
\caption{Restoration of the raw 3D image using deconvolution with the predicted 3D PSF. The central 2D slices are shown.
}
\label{restorraw}
\end{figure}

\begin{figure}[htb]
\centerline{\includegraphics[width=\linewidth]{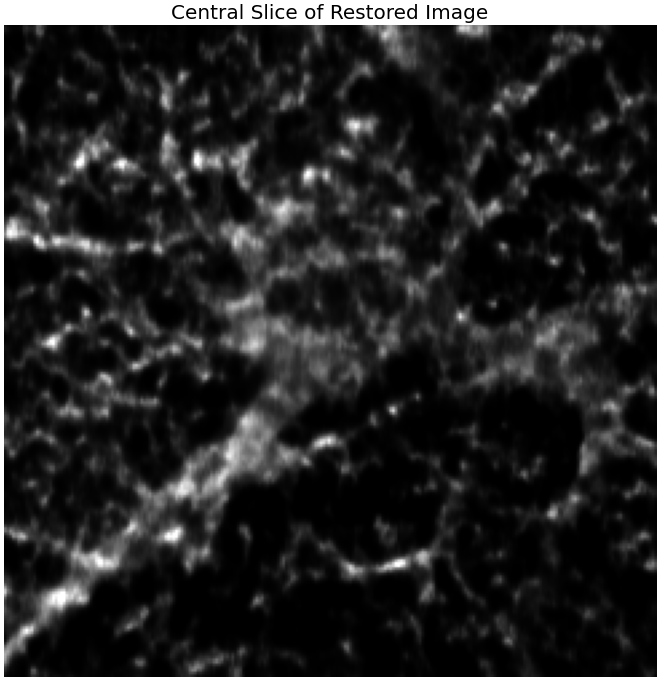}}
\caption{Restoration of the cropped source 3D image (Fig.~\ref{fig4}) using deconvolution with the predicted 3D PSF. The central 2D slice is shown.
}
\label{restorraw2}
\end{figure}

The raw source image  \cite{ly17} and its restoration are shown on Fig.~\ref{restorraw}. The restoration of the manually cropped area (cropping is shown on Fig.~\ref{fig4}) of the second source image \cite{ly22} is visualized on Fig.~\ref{restorraw2}.

\section{Discussion}
The images in this study are of objects of a complex structure (astrocytes), and crops from the images contain smaller objects varying in shape and size. Due to unavailability of such bioimages with aberrations experimentally introduced by an AO microscope, the aberrated images of astrocytes were synthetically generated.

For correction of substantially aberrated bioimages, we recommend to test our method on experimentally aberrated data (e.g. acquired by an AO microscope with a  deformable mirror or using a spatial light
modulator). Nevertheless, our approach demonstrates convincing improvements of the quality of the source raw images where a small residual PSF of the microscopic system should be present.

The results and conclusions are valid for biological samples of a structure and complexity similar to the ones of the protoplasmic astrocytes in the hippocampus of Rattus norvegicus. However, training PhaseNet without a phantom may work on approximately spherical objects or on samples that contain beads by design or spherical structures by nature.

We noticed that in the configuration file of the “single mode model” on the official Github page with the code for the PhaseNet paper \cite{saha2020practical}, the range for generating synthetic PSFs was set to (–0.075~{$\mu$}m, 0.075~{$\mu$}m), and decided to train our models in the same range. 

We used the learning rate of 0.0003 in training our models and observed that it leads to better results than with a learning rate of 0.0001 suggested in the PhaseNet publication. To determine the number of training epochs, we tried training for 10,000, 50,000, and 70,000 epochs. The decrease in the loss function between 50,000 and 60,000 epochs was insignificant; between 60,000 and 70,000, it reached a plateau.

\section{Conclusions}
This research describes a method for estimation of monochromatic aberrations on 3D biological images and image restoration by correction of aberrations via deconvolution with the estimated PSFs. 

PhaseNet trained on synthetically generated PSFs that represent only the effects of aberrations on point light sources may not be directly applicable to images of biological samples with complex shapes. Providing the generator with an unaberrated image of the sample (phantom) that represents the information about the shape of the object allows applying PhaseNet to nontrivial biological samples. 

We suggest that PhaseNet trained on synthetically aberrated images of astrocytes overfits on the shapes of the seen objects and cannot perform well on the completely unseen data. To ensure a good performance, we propose to include the information about the object's shape by supplying the generator with multiple phantoms representing the variety of shapes.

We modified the data generator from the PhaseNet study so that it can take a list of images as phantoms. The training procedure was modified in the way that the model can be trained with a user-defined validation dataset. 

Additionally, we include the guidelines for image restoration using the model's predictions. Moreover, we demonstrate that our approach is suitable for restoration of raw microscopic images without additionally introduced aberrations if the parameters of the microscope (refractive index, {$\lambda$}, NA, voxel size) are known.

\section*{Acknowledgments}

We thank Coleman W. Broaddus, Alexandr Dibrov, and Debayan Saha (MPI-CBG) for scientific discussions at different stages of the research. We thank Coleman W. Broaddus, Tom Burke (MPI-CBG), and Fabio Cunial (Broad Institute) for providing feedback on the manuscript.

\bibliographystyle{IEEEtran}
\bibliography{references}

\end{document}